\documentclass[aps,showpacs,preprint]{revtex4}
\usepackage{amsmath}
\usepackage{epsfig}

\begin{document}

\title{ Threshold $^3$He and $^3$H Transverse Electron Scattering Response Functions  }

\author{Winfried Leidemann$^{1,2}$, 
  Victor D. Efros$^{3}$,
  Giuseppina Orlandini$^{1,2}$,
  and Edward L. Tomusiak$^{4}$}
\affiliation{
  $^{1}$Dipartimento di Fisica, Universit\`a di Trento,
  I-38100 Trento, Italy\\
$^{2}$Istituto Nazionale di Fisica Nucleare, Gruppo Collegato
  di Trento, Italy\\
$^{3}$Russian Research Centre "Kurchatov Institute", 123182 Moscow,  Russia\\
$^{4}$Department of Physics and Astronomy
  University of Victoria, Victoria, BC V8P 1A1, Canada\\
}

\date{\today}

\begin{abstract}
The threshold transverse response functions $R_T(q,\omega)$ for $^3$He
and $^3$H are calculated using the AV18 nucleon-nucleon potential,
the UrbanaIX three-body force, and the Coulomb potential.
Final states are completely taken into account via the Lorentz integral
transform technique.
Consistent two-body $\pi$- and $\rho$-meson 
exchange currents as deduced using the Arenh\"ovel-Schwamb technique are included.
The convergence of the method is shown and a comparison of the corresponding MEC contribution is made to that of a consistent MEC
for the meson theoretical $r$-space BonnA potential.
The response $R_T$ is calculated 
in the threshold region at $q$=174, 324, and 487 MeV/c and compared with
available data.  The strong MEC 
contributions in the threshold region are nicely confirmed by  the data at $q$=324 and 487 MeV/c
although some differences between theoretical and experimental results remain. A comparison is also
made with other calculations, where the same theoretical input is used. The agreement is generally rather
good, but leaves also some space for further improvement.

\end{abstract}

\bigskip

\maketitle
\section{Introduction}

In a recent publication \cite{MEKLOTY} the LIT technique \cite{ELO94,ELOT04,ELOB07} was used
to compute the transverse response $R_T(q,\omega)$ for $^3$He
with a quasi-modern potential.  In that work
we employed the configuration space BonnA potential \cite{BonnRA}
together with the  Tucson-Melbourne' (TM') \cite{TM} NNN potential and the Coulomb
force. Rather detailed
numerical checks on our computational methods have been performed. As stated in \cite{MEKLOTY} the BonnA potential was chosen because,
being of boson-exchange character, meson exchange currents could be
determined relatively uniquely.
Although we had computed $R_T(q,\omega)$ for a wide range of
momentum transfers $q$ and excitation energies $\omega$ our choice
of the BonnA potential did not allow a detailed comparison with
other recent theoretical work \cite{Golak05,Deltuva1}.  That is because Golak {\it et al}
 \cite{Golak05} used the Argonne V18 (AV18) \cite{AV18} NN potential  with the
UrbanaIX (UIX) NNN potential \cite{UrbIX} while
Deltuva {\it et al} \cite{Deltuva1} employed a coupled channel CD-Bonn \cite{CDB}+$\Delta$ calculation.
Despite these differences there is a great similarity between our BonnA results
and those of \cite{Golak05,Deltuva1}. In particular the effects of meson exchange
currents appear very prominently in the near threshold region of the
response function.  Hence the purpose of the present paper is to examine the
threshold region again but this time with the AV18+UIX combination and with
the addition of the calculation of $R_T(q,\omega)$ for $^3$H as well. These
results should be directly comparable to the results of \cite{Golak05} except
for the fact that with the LIT method we are able to consistently include
the Coulomb interaction in the initial and final states. 

As indicated above meson exchange currents are
relatively straightforward to obtain when the NN potential is of
boson exchange type.  This is not the case with the AV18 potential and
thus a prescription is required for the construction of consistent
meson exchange currents. Several methods based on interpreting the isovector
part of the potential as due to an effective $\pi$ and $\rho$ exchange
have appeared in the literature \cite{Riska,Buchmann,ArenSchwamb}.
These methods are similar in principle and differ mainly in form.
Here we choose the technique of Arenh\"ovel and Schwamb \cite{ArenSchwamb}
whereas Golak {\it et al} employed the method of Riska \cite{Riska}.

The next section
provides a very brief review of the calculational technique. Full details
can be found in our previous paper \cite{MEKLOTY}. Following that we
describe our application of the method of \cite{ArenSchwamb} to construct
effective $\pi$ - and $\rho$-exchange currents for the AV18 potential.
Finally we present, discuss, and compare our results with experiment
as well as other theoretical calculaltions.

\section{The Method of Calculation}

The transverse response $R_T$ which
depends on the transverse nuclear current density operator
${\bf J}_T$ is given by
\begin{equation}
R_T(q,\omega)=\overline{\sum}_{M_0}\sum\!\!\!\!\!\!\!\!\int df\langle\Psi_0|
{\bf J}_T^\dag({\bf q},\omega)|\Psi_f\rangle\cdot\langle \Psi_f|{\bf J}_T({\bf q},\omega)|
\Psi_0\rangle\,\,\delta(E_f-E_0+q^2/(2M_T)-\omega).\label{rl}
\end{equation}
Here $M_T$ is the mass of the target nucleus, $\Psi_0$ and $\Psi_f$ denote the
ground and final states, respectively, while $E_0$ and $E_f$ are their
eigenenergies,
\begin{equation}
(h-E_0)\Psi_0=0,\qquad (h-E_f)\Psi_f=0 \ .
\end{equation}
Also in Eq.~(2) $h$ denotes the intrinsic nuclear non-relativistic Hamiltonian which
includes the kinetic energy terms, the 2N and 3N force terms.
In the present work the
2N + 3N interactions are taken as the AV18+UIX+Coulomb potentials.
By decomposing the transverse current into electric and magnetic multipoles 
the response function itself can be written as a sum of multipole components
via
\begin{equation}
\label{rcomp}
R_T(q,\omega)=
\frac{4\pi}{2J_0+1}\sum_{\lambda={\rm el,mag}}\sum_{Jj}(2J+1)(R_T)_J^{j\lambda}
\end{equation}
where
\begin{equation}
\label{mexp}
(R_T)_J^{j\lambda}=\sum\!\!\!\!\!\!\!\int\,df \langle q_{JM}^{j\lambda}|\Psi_f(J,M)\rangle
\langle\Psi_f(J,M)|q_{JM}^{j\lambda}\rangle\delta(E_f-E_0-\omega),
\end{equation}
$J$ and $M$ are the final state angular momentum and its projection,
and $|q_{JM}^{j\lambda}\rangle$ is given by
\begin{equation}
\label{q}
|q_{JM}^{j\lambda}\rangle = [{\cal T}_j^\lambda \otimes |\Psi_0(J_0)\rangle]_{JM} .
\end{equation}
In Eq.~(\ref{mexp}) $M$ is arbitrary while in Eq.~(\ref{q}) ${\cal T}_{jm}^\lambda$ 
are the standard electric ($\lambda$=el) or magnetic ($\lambda$=mag) multipole operators. 
Note that in \cite{MEKLOTY} we used two forms of the transverse electric operator,
one containing only the current operators as used here, and another called the
Siegert form which has a piece depending on the charge density operator as well.
There it was found that in the threshold region either form gave identical results.  
Therefore here we use non-Siegert form of the
transverse electric multipole operator.  Further we note that nucleon
form factors are the same as those used in \cite{MEKLOTY}, except for the neutron electric
form factor which is taken from \cite{Galster}.

The techniques we use in calculating the response have been largely
set out in \cite{MEKLOTY,ELOT04}. Briefly, the Lorentz transform of
the partial response $(R_T)_J^{j\lambda}$ is given by
\begin{equation}
\label{phi}
\Phi_{J}^{j\lambda,\alpha}(q,\sigma_R,\sigma_I)=\sum_n
\frac{(R_T)_{J}^{j\lambda,\alpha}(q,\omega_n)}
 {(\omega_n-\sigma_R)^2+\sigma_I^2}+
\int d\omega \frac{(R_T)_{J}^{j\lambda,\alpha}(q,\omega)}{(\omega-\sigma_R)^2+\sigma_I^2}.
\end{equation}
The sum in (\ref{phi}) corresponds to transitions to discrete
levels with excitation energy $\omega_n$. In our A=3 case there exists only one
discrete contribution corresponding to M1 elastic scattering.
In (\ref{phi}) the response is supplied with an additional superscript $\alpha$. It
specifies separate contributions to the response $(R_T)_{J}^{j\lambda}$ of
Eq.~(\ref{mexp}), e.g. a given $\alpha$ determines the isospin of the final state.
In addition it specifies contributions that correspond to components
of the multipole operators with different nucleon form factor dependencies
(for further details see \cite{MEKLOTY}).
These transforms are determined dynamically from
\begin{equation}
\label{dy}
\Phi_{J}^{j\lambda,\alpha}(q,\sigma_R,\sigma_I)=\langle{\tilde \psi_{JM}^{j\lambda,\alpha}
}|{\tilde \psi_{JM}^{j\lambda,\alpha}}\rangle,
\qquad
|{\tilde \psi_{JM}^{j\lambda,\alpha}}\rangle=
[h-\sigma_R+i\sigma_I]^{-1}|q^{j\lambda,\alpha}_{JM}\rangle.
\end{equation}
Once computed these $\Phi_{J}^{j\lambda,\alpha}(q,\sigma_R,\sigma_I)$ are inverted
separately to obtain $(R_T)_J^{j\lambda,\alpha}(q,\omega)$ and then $R_T^{\lambda,\alpha}$ from
\begin{equation}
\label{rcomp1}
R_T^{\lambda,\alpha}(q,\omega)=
\frac{4\pi}{2J_0+1}\sum_{Jj}(2J+1)(R_T)_J^{j\lambda,\alpha}(q,\omega).
\end{equation}

\section{One-Body and Two-Body Currents}

The one-body current in the present work consists of the non-relativistic current used
in \cite{MEKLOTY} plus all the relativistic corrections up to order
$M^{-2}$ i.e. it includes the spin and convection current terms which are of 
order  $M^{-1}$ plus all the terms of order $M^{-3}$.
We have calculated this operator from the expression for the corresponding 
single--particle matrix element
of the form $\langle{\bf p}_f|{\bf J}|{\bf p}_i\rangle$ given in
\cite{rgwt97}.

Below we give a short description of
the method of Arenh\"ovel and Schwamb \cite{ArenSchwamb} which we then use 
to obtain $\pi$- and $\rho$-exchange currents for the AV18 potential.
One begins with
the well-known forms for the static nucleon NN potentials due
to single $\pi$ or $\rho$ exchange, viz
\begin{eqnarray}
\label{derivs}
V_\pi({\bf r})\ =\ {\cal V}_\pi\,({\bf\tau}_1\cdot{\bf\tau}_2)({\bf\sigma}_1\cdot{\bf\nabla})
({\bf\sigma}_2\cdot{\bf\nabla})J_{m_\pi}(r), \\
V_\rho({\bf r})\ =\ {\cal V}_\rho\,({\bf\tau}_1\cdot{\bf\tau}_2)
({\bf\sigma}_1\times{\bf\nabla})({\bf\sigma}_2\times{\bf\nabla})J_{m_\rho}(r)
\end{eqnarray}
where
\begin{equation}
J_m(r)=\ \frac{e^{-mr}}{ r} .
\end{equation}
If there were a continuous set of $\pi$-like exchanges then $J_{m_\pi}(r)$ would be
replaced by 
\begin{equation}
{\cal V}_\pi J_{m_\pi}\ \to\ J_\pi\ =\ {\cal V}_\pi\int_0^\infty\ dm\,g_\pi (m)\,J_{m+m_\pi}(r)
\end{equation}
which can be interpreted as a superposition of other PS-mesons with mass
m+m$_\pi$ and $\pi$NN coupling constant densities of ${\cal V}_\pi g_\pi(m)$.
 Similarly for the vector bosons one writes
\begin{equation}
{\cal V}_\rho J_{m_\rho}\ \to\ J_\rho\ =\ {\cal V}_\pi \int_0^\infty\ dm\,g_\rho (m)\,J_{m+m_\pi}(r)
\end{equation}
where again the hypothetical vector mesons of mass $m+m_\pi$ have $\rho$NN coupling
constant densities of ${\cal V}_\pi g_\rho(m)$.
Note that the masses of the hypothetical vector bosons here extend from
m$_\pi$ to $\infty$. Application of the derivatives in Eq.~(\ref{derivs}) gives the
central and tensor potentials as
\begin{eqnarray}
\label{vcvt}
V^C(r)=\ \frac{1}{3}{\cal V}_\pi\,J_{m_\pi}(r)\,\int_0^\infty\ dm\,[ g_\pi (m)\,+\,2 g_\rho (m)]
\,(m+m_\pi)^2\,e^{-mr} \nonumber \\
V^T(r)= \frac{1}{3}{\cal V}_\pi\,J_{m_\pi}(r)\,\int_0^\infty\ dm\,[ g_\pi (m)\,-\, g_\rho (m)]
\,(m+m_\pi)^2\,F_T[(m+m_\pi)r]\,e^{-mr}, \nonumber\\
\end{eqnarray}
where 
\begin{equation}
F_T(mr)\ =\ 1\,+\,\frac{3}{mr}\left(1\,+\,\frac{1}{mr}\right) \nonumber
\end{equation}
and where a delta function $\delta({\bf r})$ is removed from $V^C(r)$ by imposing the condition
\begin{equation}
\label{delrm}
\int_0^\infty\ dm\,[ g_\pi (m)\,+\,2 g_\rho (m)]\ =\ 0 .
\end{equation}
 The change of variable
\begin{equation}
m\ =\ s\,\tan\left[\frac{\pi}{4}(x + 1)\right]
\end{equation}
and the introduction of an $N$-point Gaussian integration in $x$=[-1,1] gives
\begin{equation}
\int_0^\infty\,f(m)\,dm\ \to\ \sum_{j=1}^N\,\bar w_j\,f(m(x_j))
\end{equation}
where 
\begin{equation}
\bar w_j\ =\ s\frac{\pi}{4}\,\sec^2\left[\frac{\pi}{4}(x_j+1)\right]\,w_j \ .
\end{equation}
Here the $x_j$ and $w_j$ are the $N$ abscissae and weights for the quadrature
integration and $s$ is a parameter to be discussed shortly.  If $N$ values of $r$ are selected,
say $r_i$ $i=1...N$, then Eqs.~(\ref{vcvt}) can be solved for $g_\pi(m_j$) and $g_\rho(m_j$)
$j=1...N$.  We follow Ref.~\cite{ArenSchwamb} in using their method to choose $N$ values of $r$
between $r_{min}$ = 0.01 fm and $r_{max}$=12 fm .  These values of the $g_\pi$ and $g_\rho$
thus give an exact fit to the AV18 values of $V^C$(r) and $V^T$(r) at the chosen
$N$ values of $r$.  The parameter $s$ is adjusted to give the minimal absolute deviation between
the original potential and its new parametrization when integrating the absolute deviation from $r_{min}$ to $r_{max}$. 
Up to this
stage the $\delta$-function removing condition Eq.~(\ref{delrm}) has not yet been imposed. Ref.~\cite{ArenSchwamb}
imposes this condition by modifying the values of g$_\pi(N)$ and g$_\rho(N)$ i.e.
the contributions of the largest mass terms such that Eq.~({\ref{delrm}) is numerically
satisfied.  Thus if the least squares fitting yields parameter values such that
\begin{equation}
\int\,dm [g_\pi(m)\,+\,2g_\rho(m)]\ \to\ \sum_j\ \bar w_j [g_\pi(m_j)\,+\,2g_\rho(m_j)]\ =\ C\neq 0
\end{equation}
then the replacements
\begin{equation}
g_\pi(N)\ \to\ g_\pi^{Old}(N) - \frac{C}{3 \bar w_N}\quad\hbox{and}\quad
g_\rho(N)\ =\ g_\rho^{Old}(N) - \frac{C}{3 \bar w_N}
\end{equation}
will satisfy Eq.~(\ref{delrm}) at the expense of spoiling the short range fit at $r$=0.01 fm.
However the fit is still very good for $r$ greater than approximately 0.06 fm depending on
the choice of $N$. 

Finally the $\pi$- and $\rho$-meson exchange current multipoles obtained from the above
method are, apart from slight modifications, equal to those
listed in Appendix C of \cite{MEKLOTY}.  These modifications in the case of $\pi$-exchange currents
are the following: (a) here we use $f_0^2$=0.075 and $m_\pi$=0.70 fm$^{-1}$ , (b) $H^\pi(r)=
\sum_{j=1}^N\,\bar w_j\,g_\pi(m_j)\,J_{m_\pi + m_j}(r)$, (c) 
 $\Phi_{\sigma,\ell}^{(n)}(q,r)=\sum_{j=1}^N  \bar w_j g_\pi(m_j)\,\phi_{\sigma,\ell}^{(n)}(q,r,m_j)$,
where the functions $\phi_{\sigma,\ell}^{(n)}(q,r,m)$ are defined in \cite{fa}. The multipoles
of the $\rho$-exchange currrents are obtained from the $\pi$-meson exchange
currents  by the replacements:
$H^\pi(r) \to H^\rho(r)$, $g_\pi(m_j) \to$ $g_\rho(m_j)$ and by inserting into each equation the factor
\begin{equation}
-6 (-1)^\rho \left\{\begin{array}{ccc}1&1&1\\ 1&\rho&1\end{array}\right\}.
\end{equation}

\section{Results and Discussion} 

For the calculation of $R_T$ we take into account electric and magnetic
multipole transitions to final states with total angular momentum $J_f$
up to 5/2 and in the case of final isospin $T_f=1/2$ we include transitions 
to the $J_f$=7/2 state in addition. The latter lead to a small contribution in the peak
region for $q$=174 MeV/c (for this rather small $q$ the "quasielastic" peak 
is a part of the threshold region). The three-nucleon ground states and the
various LITs are calculated using expansions in correlated hyperspherical harmonics 
\cite{ELOB07}. The expansions in hyperradial and hyperspherical basis functions
is made such that the convergence errors of the various LITs are considerably
smaller than 0.5\% in the energy region of main interest (-10 MeV $< \sigma_R < 20$ MeV). 
As the resolution parameter $\sigma_I$ we choose 10 MeV, but for the two dominant 
threshold contributions (M1 and M2 transitions to $T_f$=1/2 states with $J_f=1/2$
and 5/2, respectively) we take $\sigma_I=2.5$ MeV extending the HH expansion further in order 
to guarantee also in these cases the requested small convergence error. Before carrying out 
the inversion of the LIT we first subtract the elastic M1 contribution. The inversion itself 
proceeds in the following way. The above mentioned M1 and M2 transitions are inverted separately. 
Because of their different break-up thresholds the LITs of the remaining multipole contributions are 
individually summed up and inverted for the $T_f=1/2$ and 3/2 states. As to the inversion method we use 
our standard expansion of the LIT over given basis functions, where also 
the threshold behavior is incorporated (see \cite{ELO99,Andreasi,ELOB07}). 
Note that in case of $^3$He we also take into account 
the effect of the Coulomb barrier on the threshold behavior.

In the first part of the discussion we consider some aspects of the consistent MEC for the AV18 potential. In Fig.~1 we show for $^3$He the LIT of the MEC contribution for the M1 and E1 transitions to the $J_f=1/2$, $T_f=1/2$ final state at $q$=174 MeV/c for three values of the Gaussian integration parameter $N$ discussed in section III. 
One sees that $N=12$ does not yet lead to a convergent result, but that the $N=14$ and $N=16$ results are almost identical. The convergence of other analyzed MEC transitions looks very similar. Thus, in order to be on the safe side we use $N=16$ for the calculation of the MEC contribution of $R_T$. 

Next we make a comparison of the consistent MEC between BonnA and AV18 potentials (as mentioned before the unique BonnA-MEC was calculated by us in \cite{MEKLOTY}). The MEC effect on the threshold response is much stronger for the magnetic than for the electric transition. Therefore we investigate the magnetic MEC contribution considering the LIT of the total MEC magnetic transition strength (sum of LITs of all MEC magnetic multipole transitions).
In Fig.~2 we show the $^3$He results for the two $T_f$ channels separately at $q$=324 
MeV/c.
One sees that the shape of the LITs for BonnA and AV18 MEC strength is very
similar, in particular for the $T_f=1/2$ channel. In fact the main difference consists in the overall strength. In comparison to the BonnA case the AV18 MEC strength is reduced by about 12\% ($T_f=1/2$) and 9\% ($T_f=3/2$). For the other two $q$ values, not shown in the figure, the shapes are also very similar, but here one finds somewhat different reductions of the overall strength, namely 9\% and 15\% ($T_f=1/2$) and 4\% and 13\% ($T_f=3/2$) at $q=174$ and 487 MeV/c, respectively. These results show that the ratio of the BonnA to AV18 MEC strength grows with increasing momentum transfer. 

Now we turn our attention to the comparison of our $R_T$ results to the data
of Retzlaff {\it et al} \cite{Retzlaff} (see Fig.~3). In case of $^3$He one has a rather good agreement of theoretical and experimental transverse response functions for the two higher $q$ values. The MEC contribution is essential for reaching this agreement. At $q$=174 MeV, however, the theoretical $R_T$ underestimates the data below 11 MeV. In the triton case the situation looks worse. Already for the two higher $q$ values one finds a slight underestimation of the data, in addition the discrepancy becomes even larger at the lowest $q$. The relativistic contributions originating from the use of a relativistic one-body current are not negligible already at $q$=323 MeV/c and make the discrepancy theory-experiment somewhat larger. One can conclude that the present agreement between theory and experiment is not bad, but certainly not very good. It seems that a different nuclear force does not improve the situation, since our $^3$He results at $q$=174 MeV/c with the BonnA+TM' potential from \cite{MEKLOTY} is almost identical to the AV18+UIX result (the $^3$H case was not considered in \cite{MEKLOTY}). Additional currents involving the $\Delta$ resonance, up to now only partially considered in the literature for the threshold kinematics (see \cite{Pisa2000,Deltuva1}), could probably lead to a small improvement. On the other hand, as to the experimental cross sections below the break-up threshold, one cannot exclude that the data are systematically a little bit too high. Thus, presently, one cannot speak of a serious disagreement of theoretical and experimental results.

Finally we want to make a brief comparison of our results for the threshold response with other calculations with the same theoretical input. As already mentioned above there is the calculation of \cite{Golak05}, where a nonrelativistic one-body current and a consistent AV18-MEC have been taken as the current operator. For the small energy range up to the three-body break-up threshold there exists in case of $^3$He another calculation \cite{Pisa2000}. Here we only consider the results with a nonrelativistic one-body current, since the MEC of \cite{Pisa2000} includes also an additional current involving the $\Delta$ resonance, which makes a consistent comparison more difficult. In Figs.~4 and 5 we illustrate the various results (note no results available from \cite{Golak05} with one-body current only). For the $R_T$ of $^3$H one has quite a good agreement between
the different theoretical calculations. Some differences are visible below the three-body break-up threshold and above 14 MeV,
only for the highest $q$ value one finds also some small differences for other energies. For the latter a not unimportant part of the
difference can be explained by the use of the three-momentum transfer squared for the nucleon form factors in case of \cite{Golak05}, while we take the four-momentum transfer squared. This leads to a relative reduction of the $R_T$ of \cite{Golak05} by a little bit more than 1\% (note that at $q$=487 MeV one has a nuclear recoil energy of about 50 MeV). For the $R_T$ of $^3$He we first consider the result of \cite{Pisa2000} (see Fig.~5). Up to about 1 MeV above threshold our results are about 20\% smaller. Close to the three-body break-up threshold differences are reduced to 9\% ($q$=174 MeV/c), 3\% ($q$=323 MeV/c) and 0\% ($q$=487 MeV/c). Here we should mention that for the longitudinal response function $R_L$ a much better agreement between the two calculations was obtained \cite{ELOT04}. For the $^3$He results in Fig.~4 one has to take into account that there is no Coulomb force in the final state interaction of \cite{Golak05}. This should explain why differences with our results are considerably larger than for the $^3$H case, even though the effects seem to be a bit too large in order to be caused by the Coulomb force alone.

Summing up we can say the following. For the AV18+UIX nuclear interaction we have calculated the ($e,e'$) transverse response function $R_T(q,\omega)$ of $^3$H and $^3$He close to the break-up threshold at three momentum transfers ($q$=174, 323, and 487 MeV/c). Besides the one-body current a consistent isovector MEC for the AV18 potential has been employed. The consistent MEC has been constructed using the Arenh\"ovel-Schwamb method, where an expansion over fictitious meson masses is made. The convergence of the expansion has been shown for our results. In comparison to the MEC contribution of the meson theoretical BonnA potential one obtains very similar results for the energy dependence, but the absolute size differs somewhat and is function of $q$ and the final isospin channel. Relativistic effects for the one-body current have been included and lead to non negligible contributions for the two higher $q$ values. A comparison with experimental data has been made and, as already known before in the literature, MEC contributions lead to a much improved agreement with data. As discussed in the text the comparison
theory-experiment is not bad, but certainly not yet completely satisfying. The agreement with other theoretical calculations
with the same theoretical input is overall rather good, but some differences, in particular below the three-body break-up threshold, are also evident.

\section{Acknowledgment}

We thank J. Golak, R. Skibinski, and M. Viviani for providing us with their theoretical results.
Acknowledgements of financial support are given to
the RFBR, grant 07-02-01222-a and RMES, grant NS-3004.2008.2 (V.D.E.),
and to the National Science and Engineering Research Council of Canada (E.L.T.).

\newpage

\centerline{CAPTIONS TO FIGURES}

\begin{figure}[ht]
\centerline{\resizebox*{16cm}{11cm}{\includegraphics*[angle=0]{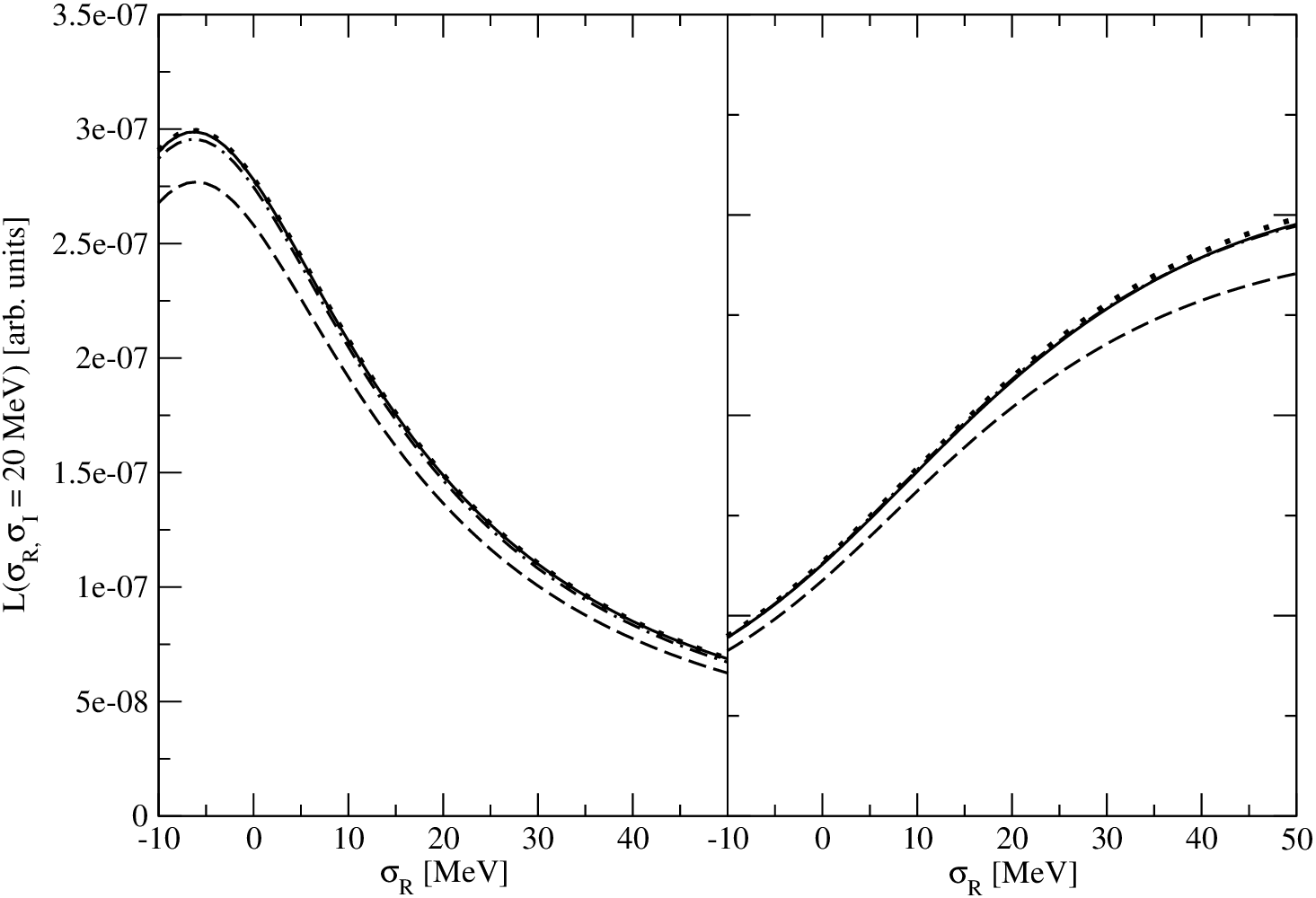}}}
\caption{Lorentz integral transforms of the M1 MEC transition strength to the final state $J^\pi$= 1$^+$/2,
$T_f$=1/2 (left panel) and of the E1 MEC transition strength to the final state $J^\pi$= 1$^-$/2, $T_f$=1/2 (right panel)
calculated with a different number $N$ of hypothetical meson masses (see text): $N$=10 (dashed), $N$=12 (dash-dotted),
$N$=14 (solid), $N$=16 (dotted).}
\label{fig1}
\end{figure}

\begin{figure}[ht]
\centerline{\resizebox*{16cm}{11cm}{\includegraphics*[angle=0]{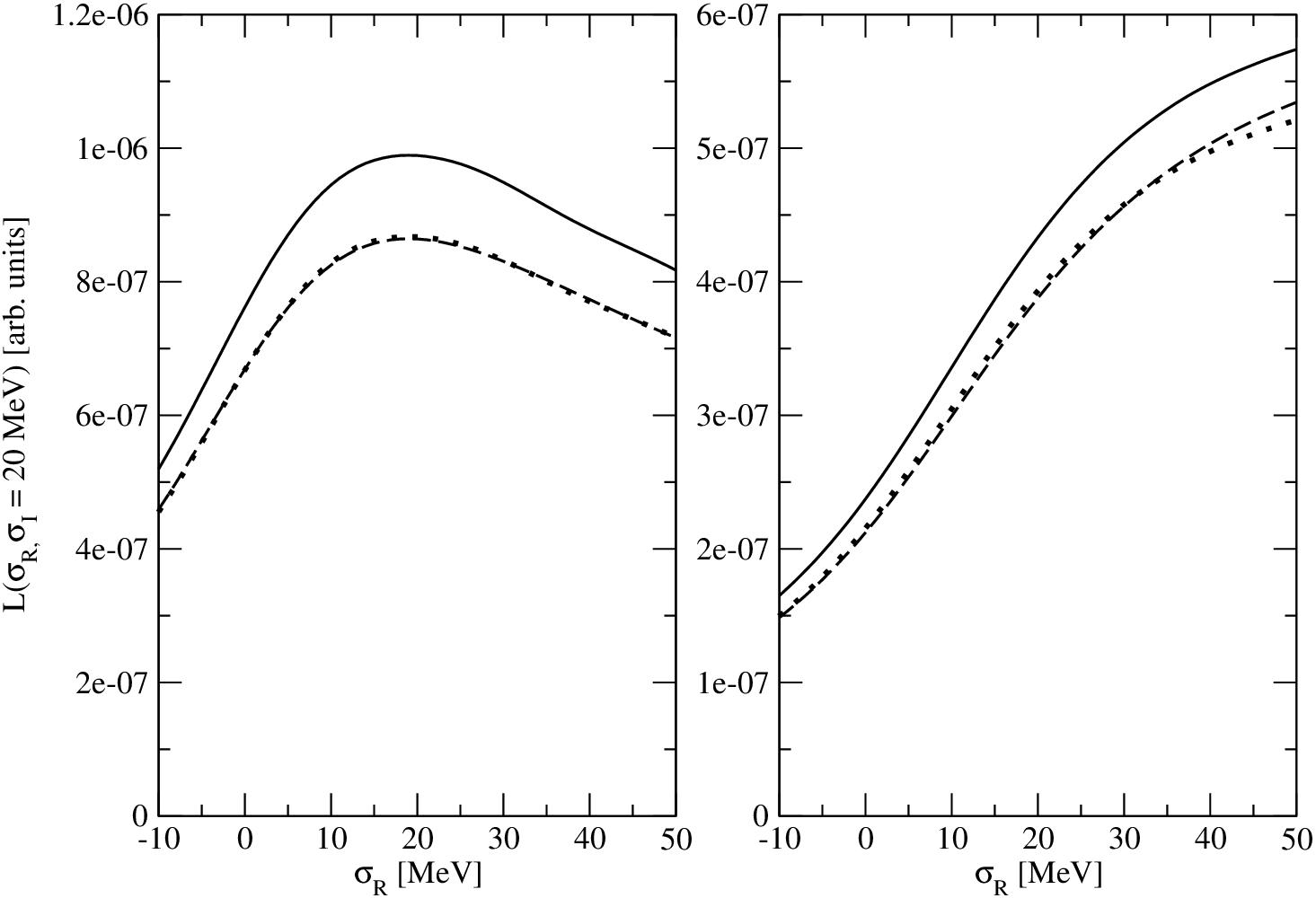}}}
\caption{Lorentz integral transform of the total inelastic magnetic MEC contribution for final isospin channel
$T_f$=1/2 (left panel) and $T_f$=3/2 (right panel): consistent $\pi$ and $\rho$ MEC for AV18 potential
evaluated for $^3$He with AV18+UIX potentials (dashed), consistent $\pi$ and $\rho$ MEC for BonnA potential
evaluated for $^3$He with BonnA+TM' potentials (solid), latter result renormalized with factors
0.88 ($T_f$=1/2) and 0.91 ($T_f$=3/2) (dotted).}
\label{fig2}
\end{figure}

\begin{figure}[ht]
\centerline{\resizebox*{16cm}{11cm}{\includegraphics*[angle=0]{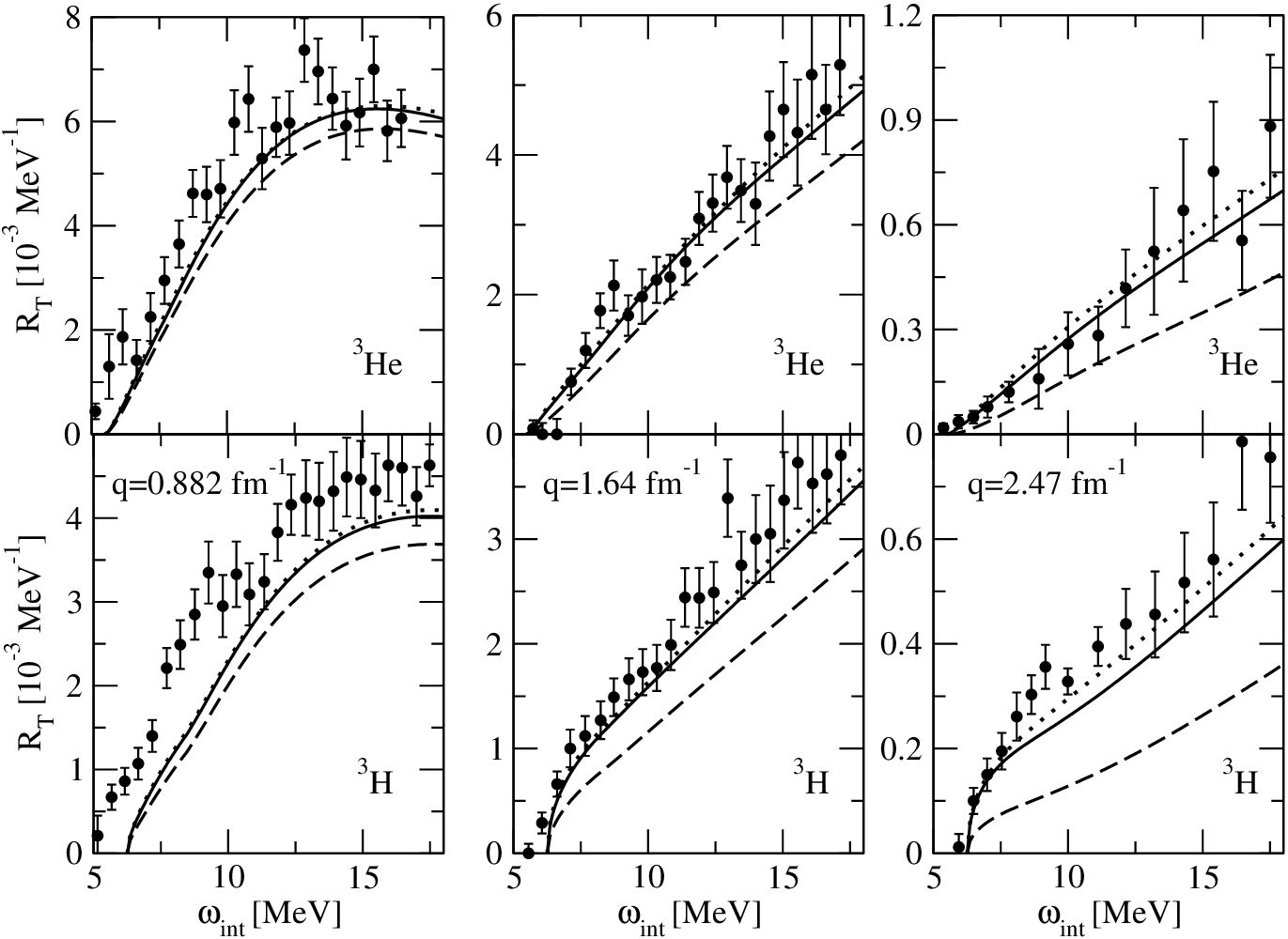}}}
\caption{Comparison of experimental and theoretical results for the threshold $^3$He (upper panels)
and $^3$H (lower panels) transverse response function $R_T$ as function of internal excitation energy $\omega_{\rm int}$ at three momentum transfers $q$:
174 MeV/c (left panels), 324 MeV/c (middle panels), 487 MeV/c (right panels). Theoretical results with
different current operators: relativistic one-body current (dashed), relativistic one-body current + MEC
(solid), nonrelativistic one-body current + MEC (dotted). Experimental data from \cite{Retzlaff}.}
\label{fig3}
\end{figure}

\begin{figure}[ht]
\centerline{\resizebox*{16cm}{11cm}{\includegraphics*[angle=0]{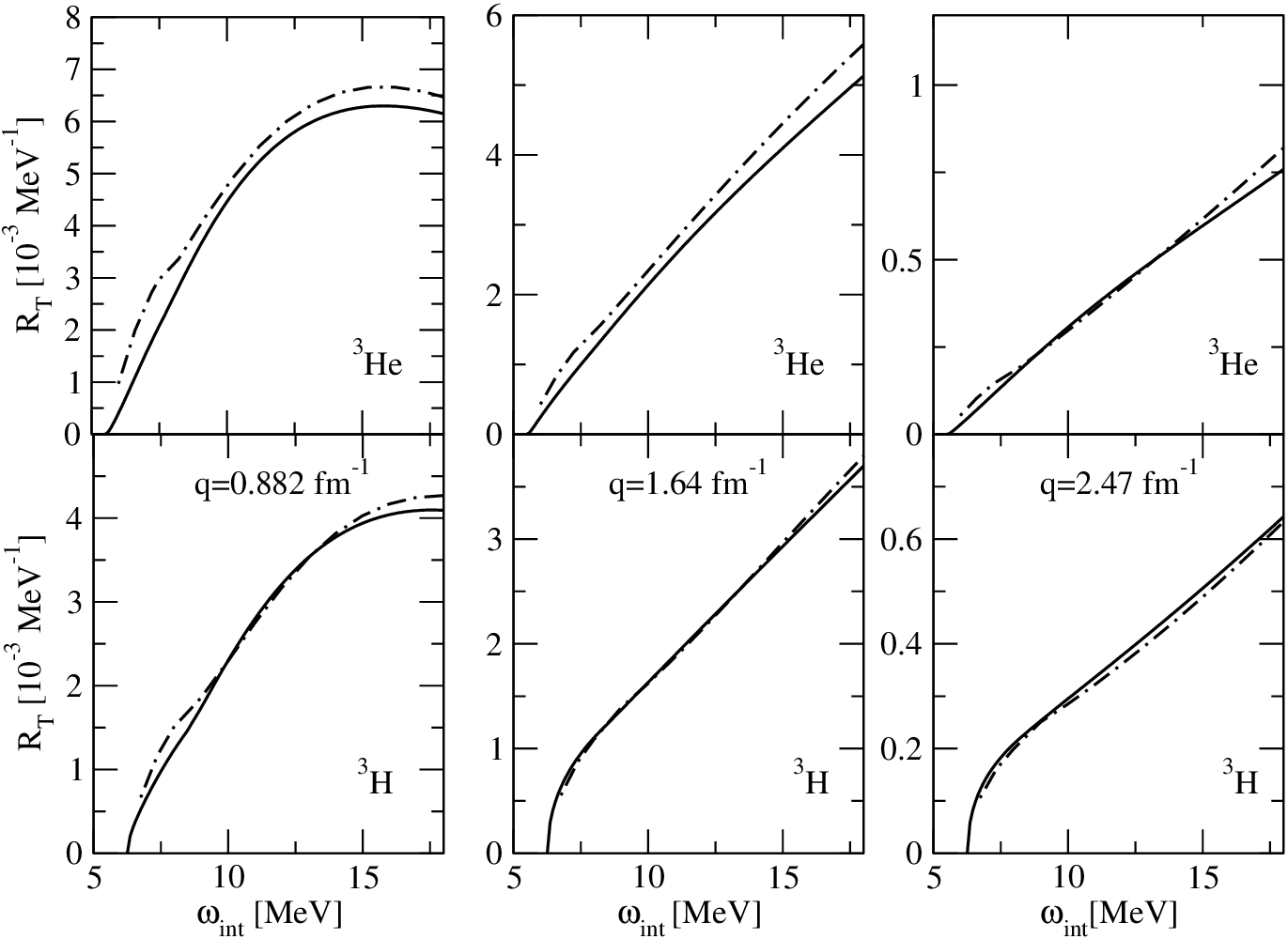}}}
\caption{As Fig.~3, but for comparison of different theoretical calculations. $R_T$ with nonrelativistic one-body 
current + MEC from present work (solid) and from \cite{Golak05} (dash-dotted).}
\label{fig4}
\end{figure}

\begin{figure}[ht]
\centerline{\resizebox*{16cm}{11cm}{\includegraphics*[angle=0]{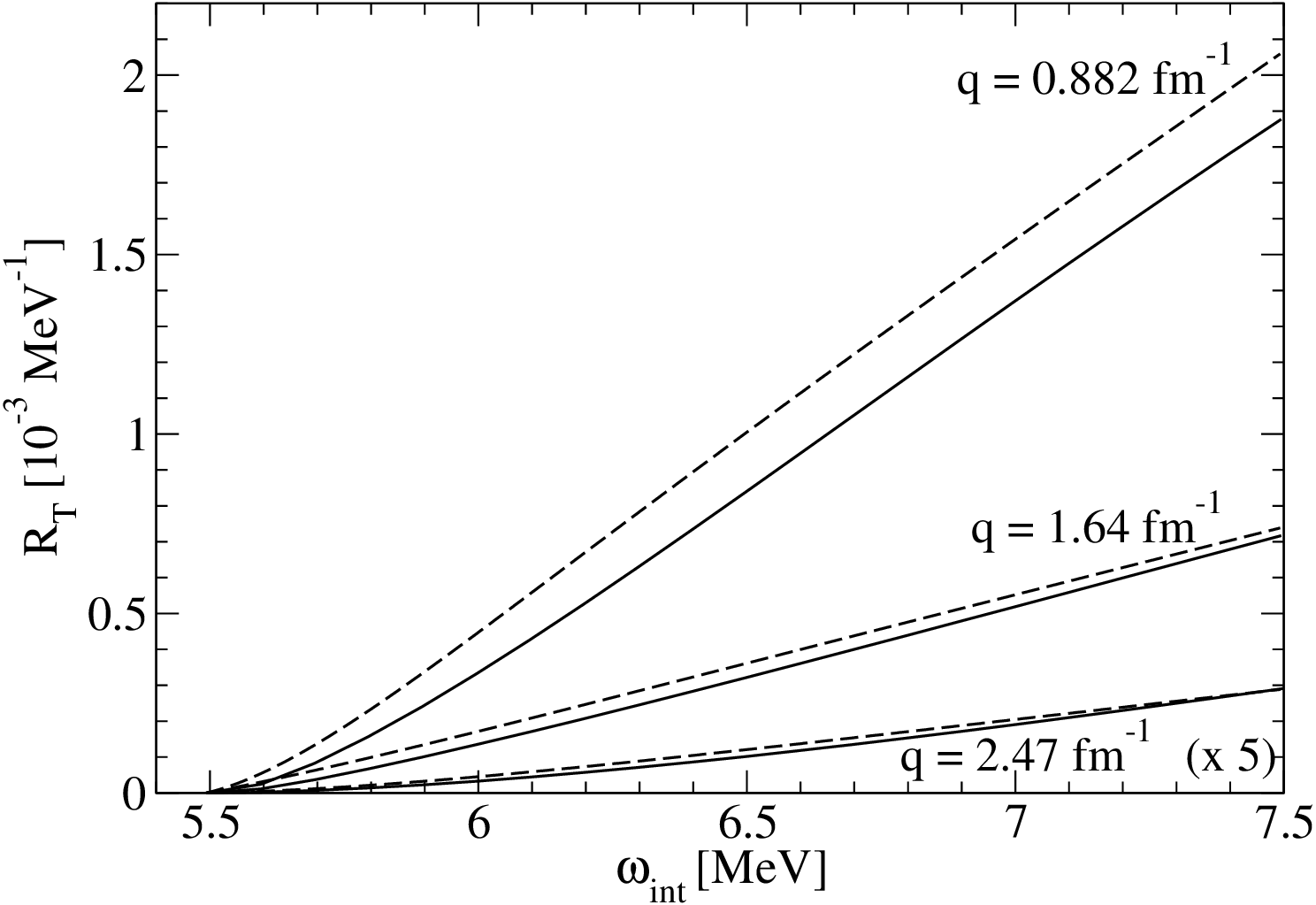}}}
\caption{Comparison of $R_T$ with nonrelativistic one-body current
from present work (solid) and from \cite{Pisa2000} (dashed) at various $q$ as indicated in figure (note that the
result at $q$=2.47 fm$^{-1}$ is multiplied by a factor of 5).}
\label{fig5}
\end{figure}

\end{document}